\documentclass[epj]{svjour}
\usepackage{graphics,amsmath}

\def\(({\left(}
\def\)){\right)}
\def\[[{\left[}
\def\]]{\right]}

\usepackage{graphics,epsfig}
\begin{document}
\title{Chaotic temperature dependence in a model of spin glasses}
\subtitle{A Random-Energy Random-Entropy model}

\author{F.~Krzakala\inst{1} \and O.C.~Martin\inst{1,2}}
\offprints{krzakala@ipno.in2p3.fr}
\institute{Laboratoire de Physique Th\'eorique et Mod\`eles Statistiques,
b\^at. 100, Universit\'e Paris-Sud, F--91405 Orsay, France
\and Service de Physique de l'\'Etat Condens\'e,
Orme des Merisiers --- CEA Saclay, 91191 Gif sur Yvette Cedex, France.}

\date{\today}

\abstract{
We address the problem of chaotic temperature dependence 
in disordered glassy systems at equilibrium
by following states of a random-energy random-entropy model 
in temperature; of particular interest are the crossings of the
free-energies of these states. We 
find that this model exhibits strong, weak or no temperature chaos
depending on the value of an exponent. This allows us to
write a general criterion for temperature chaos in disordered systems,
predicting the presence of
temperature chaos in the Sherrington-Kirkpatrick and
Edwards-Anderson spin glass models, albeit when the number of
spins is large enough. The absence of chaos for
smaller systems may justify why it is difficult
to observe chaos with current simulations. We also 
illustrate our findings by studying 
temperature chaos in the na\"{\i}ve mean field equations 
for the Edwards-Anderson spin glass.}

\maketitle
%

\section{Introduction}
\label{sect_intro}

It is generally agreed that the equilibrium states of 
spin glasses~\cite{MezardParisi87b,Young98} are ``fragile'', i.e.,
they are sensitive even to very small perturbations. Roughly,
this can be argued from the fact that there exist many (meta-stable)
states whose excess free-energies grow more slowly than the
system's volume; then arbitrarily small extensive
perturbations will re-shuffle the different (meta-stable) states
and the lowest free-energy state will be completely different from
the one without the perturbation. This phenomenon is referred
to as ``chaos''~\cite{BrayMoore87} as the equilibrium state 
depends chaotically on the perturbation.
In particular, there has long been a consensus that chaos arises 
when changing the couplings $J_{ij}$ between the spins. However, 
the situation is quite different when the parameter being changed
is the temperature; though earlier work~\cite{Kondor83,KondorVegso93}
claimed that there is chaos under even 
infinitesimal temperature changes, some recent
work goes against this; indeed, analytic expansions~\cite{Rizzo01}
around
the critical temperature give no temperature chaos in the 
Sherrington-Kirkpatrick (SK) model, and extensive Monte Carlo simulations 
in both the three-dimensional Edwards-Anderson (EA) model and the SK model
find little or no evidence of 
temperature chaos~\cite{BilloireMarinari00,BilloireMarinari02}.

This controversy is interesting in itself as it shows just how poorly
understood spin glasses remain after years of work. 
More importantly perhaps, temperature chaos plays a central role
in many theoretical descriptions of dynamics. Indeed, 
phenomena such as rejuvenation~\cite{JonasonVincent98}
that follow temperature changes in glassy systems must arise if 
there is temperature chaos
(see for instance~\cite{BouchaudCugliandolo98} and~\cite{YoshinoLemaitre00}).
Thus both for equilibrium and for non-equilibrium properties, 
it is appropriate to understand whether or not there is temperature chaos 
in disordered systems, and in spin glasses in particular.

In this work, we present a solvable glassy model
where the temperature dependency is easily analyzed. Our main result
is that the presence of temperature chaos depends on an exponent
associated with state-to-state entropy fluctuations. 
Depending on the value of this exponent, 
the model exhibits strong, weak, or no chaos at all. When 
extrapolating these results 
to realistic models, we find chaos both for the SK and EA models but
only when the number of spins $N$ is quite large, $O(1000)$.
This may justify why chaos
is not seen in Monte-Carlo simulations. Finally, we attempt to confirm
numerically these predictions by using a na\"{\i}ve mean field
approximation; there we
also find the expected signatures of a chaotic temperature dependence.

This paper is organized as follows. In section~\ref{sect_general}, 
we motivate and explain this study; our Random-Energy Random-Entropy Model 
is described and solved in 
section~\ref{sect_rerem}. In section~\ref{sect_cross}, we 
determine how the equilibrium state changes with temperature.
In section~\ref{sect_droplet}, we generalize this model to the droplet/scaling
picture. In section~\ref{sect_chaos}, we discuss the predictions coming
from another simple spin glass system based on the 
na\"{\i}ve mean field equations. 
We summarize and conclude in section~\ref{sect_discussion}.

\section{Temperature chaos as level crossings}
\label{sect_general}

Our framework is motivated by the concept of {\it ``valleys''} in glassy
systems. Indeed, within the mean field picture of disordered systems in
their low temperature frozen phase, ergodicity is broken and
configuration space is broken into many components not related by
symmetry. These components are often referred to as valleys as in
an energy landscape picture. Loosely, one
considers each valley to be associated with one thermodynamic ``state'',
though such an association has ambiguities, especially in
finite volume. This picture of valleys is {\it not} restricted
to a mean field framework: even in the droplet~\cite{FisherHuse86} 
and scaling~\cite{BrayMoore86} pictures, the energy landscape is 
very rugged and
there exist multiple valleys corresponding to 
possibly (meta-stable) states that 
do not contribute to equilibrium properties in the thermodynamic limit.
We will thus assume that we can talk of states (equilibrium
or meta-stable) and then ask within such a picture what
can be said about temperature chaos.

\subsection{Level crossings}

In the simple case of the ferromagnetic Ising model below its 
critical temperature $T_c$, there are two valleys 
${\cal V}_1$ and ${\cal V}_2$ with 
degenerate free-energies and magnetizations
$\pm m(T)$. Consider
the mean spin-spin overlap $q^{1,2}$
between these valleys: it is $q^{1,2}=- m^2(T)$.
The associated (equilibrium) states have a smooth evolution 
with temperature, merging
when $T$ approaches $T_c$ from below. It is also possible to consider
the overlaps between
an equilibrium state at $T$ and one at $T+\delta T$. Such an overlap
is equal to $\pm m(T) m(T+\delta T)$, which also varies smoothly with
$T$ and $\delta T$. On the contrary, to have temperature chaos, 
we want such overlaps to be very sensitive to $\delta T$, and
generically one expects the equilibrium states
at different temperatures to be ``totally different'', meaning their
overlap is zero (in the absence of an external field).

\begin{figure}
\centerline{\hbox{\epsfig{figure=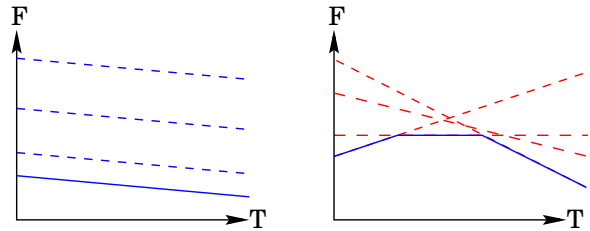,width=8cm}}}
\caption{Temperature dependence of free-energies of the lowest 
states. (a): there are no entropic fluctuations, as in 
the spherical p-spin model; (b) possible level crossing
behavior in more realistic models.}
\label{fig_levelcross}
\end{figure}

Does one have temperature chaos in disordered systems where 
the abundance of (meta-stable) states plays a fundamental role?
The answer is yes for spin glasses on Migdal-Kadanoff
hierarchical lattices~\cite{BanavarBray87,NifleHilhorst92} or
for other disordered systems like the Directed Polymer 
in a Random Medium in 1+1 dimension~\cite{FisherHuse91,SalesYoshino02a}. 
But in other systems one might instead expect that
a state dominating the partition function at $T$
will also dominate it at $T+\delta T$. Such a scenario is represented
pictorially on the left of figure~\ref{fig_levelcross} where
for each state we plot its free-energy as a function of temperature.
We can think of these curves as a family of levels, and in
this case there are no level crossings. 
Interestingly, this is what occurs in the
the infinite range {\it spherical} p-spin 
model~\cite{BarratFranz97}. 
There one can define states in the large volume limit, 
for instance by the Thouless-Anderson-Palmer 
(TAP)~\cite{ThoulessAnderson77} equations.
These different states
evolve smoothly with temperature, and furthermore 
have zero mutual overlap. When
following analytically the free-energy of these states as a function of
temperature, one finds that they keep the same order:
there are {\it no level crossings}.

Although such a property may be special to the spherical p-spin model,
there is also the possibility that it is more general, extending
to all mean field models or even to finite 
dimensional models; this is supported by the lack of any
numerical evidence in favor of chaos in either the SK model or
the Edwards-Anderson model.

\subsection{The Bray and Moore argument }

To understand why many researchers expect 
a chaotic temperature dependence
in finite dimensional spin glasses, it is of 
some use to go over the argument due to Bray 
and Moore (BM)~\cite{BrayMoore87},
reformulated for our particular purposes. For specificity, we work 
with the $3$-dimensional EA model with no magnetic 
field whose Hamiltonian is

\begin{equation}
H = - \sum_{<ij>} J_{ij} S_i S_j
\label{Hamil}
\end{equation}

The $S_i$ are Ising spins on an $L \times L \times L$
lattice and the sum is over nearest neighbor sites with
periodic boundary conditions; the $J_{ij}$ are independent Gaussian 
random variables. We start from two ``states'' at low temperature,
differing by a very large droplet and work in the framework of
the droplet/scaling theory of spin glasses~\cite{BrayMoore86,FisherHuse88}.
The two states have free-energies that differ 
by $\Delta F(T) = \Delta E - T \Delta S \approx \Upsilon {\ell}^{\theta}$, 
where $\ell$ is the characteristic size of the flipped droplet 
when going from state 1 to state 2 and $\Upsilon$ is the temperature
dependent stiffness coefficient. Now change the 
temperature; $\Delta F(T+\delta T) \approx \Delta E - (T+\delta T) \Delta S$, 
so that 
$\Delta F(T+\delta T) \approx \Upsilon {\ell}^{\theta} - \delta T \Delta S $.

Following Bray and Moore, it seems inevitable that the entropy difference 
is associated with the droplet's surface so that 
$\Delta S$ has a random sign and a typical magnitude proportional
to $\ell^{d_s/2}$ where $d_s$ is the fractal dimension of
the droplet's surface. Then 
if ${d_s}/{2}>{\theta}$, which follows from 
the inequality $\theta < (d-1)/2$, the difference in free-energies 
$\Delta F(T+\delta T) \propto \Upsilon {\ell}^{\theta} 
+ \delta T {\ell}^{d_s/2}$ 
can change sign between $T$ and $T+\delta T$. 
The conclusion is then that
the equilibrium state(s) between $T$ and $T+\delta T$
should change on a length scale $\ell$ scaling as
$\ell(\delta T) \propto \delta T^{-1/(\frac{d_s}{2}-\theta)}$ while
their typical overlap
$q_{T,T+\delta T}$
should go to 
zero as
\begin{equation}
\label{eq_exp_decay}
q_{T,T+\delta T} \approx {\(( \frac{l \(( \delta T \)) }{L}  \))}^{d/2}
\end{equation}
where $L$ is the total lattice size. In the droplet/scaling
picture, $\theta\approx 0.2$, while in the mean field picture
$\theta = 0$; in either case, the argument strongly suggests
that the EA model has temperature chaos.
At the root of the argument are large cancellations between $E$ and $S$
in the equilibrium value of $F$; these cancellations make
the equilibrium state very sensitive to changes in $T$.
Note that numerical evidence for such cancellations
has been found~\cite{MarinariMartin01} in the $3-d$ EA model.
Finally, as mentioned previously,
chaos does occur on the Migdal-Kadanoff lattice;

To us, the Bray and Moore argument is very convincing 
if one goes from $T=0$ to $T=\epsilon$
for any $\epsilon > 0$, suggesting that chaos is unavoidable
at $T=0$. However, the argument has potential pitfalls
when considering chaos at $T>0$. First, when going from 
$T$ to $T+\delta T$, both $\Delta E$ and $\Delta S$ change, 
a fact which is neglected in the argument. Second, one cannot
treat a single scale $\ell$ alone; it is possible that the {\it droplets}
on the scale smaller than $\ell$ will see their free-energy change
sign, and thus the ``evolution'' of ${\cal V}_1(T)$ to $T+\delta T$
will differ from ${\cal V}_1(T)$ by many droplet excitations.
This evolution may be enough
to keep the free-energy of ${\cal V}_1(T+\delta T)$ below that
of ${\cal V}_2(T+\delta T)$. In this sense, the chaos at
scale $\ell(\delta T)$ might be quenched by the way
the equilibrium state {\it adapts} to changes in $T$. And since
${\cal V}_1(T)$ and ${\cal V}_1(T+\delta T)$ 
differ only by droplet excitations, their overlap will 
not be close to zero. Unfortunately we cannot take into account these
effects in a realistic way; nevertheless, we want to use
the fact that the state dominating the partition function
at one temperature is quite special, being
the result of an extreme statistic where the free-energy
is optimized for that temperature. We now proceed to
see how we can model this.

\section{The Random-Energy Random-Entropy Model}
\label{sect_rerem}

In the Bray and Moore argument, the important feature is that there are 
very large
entropy fluctuations from state to state. We shall now define a glass model 
that incorporates such entropy fluctuations. 
Since our model is close to
Derrida's Random Energy Model~\cite{Derrida81}(REM),
 we have named it the {\it Random-energy 
Random-entropy model} (Re-Rem).
In the spirit of the REM, we
start by taking $2^N$ ``valleys'', each
 with an energy taken from a Gaussian 
distribution, but we shall also assign a random entropy
to each valley.

\subsection{Definition of the Re-Rem }

The Re-Rem model is defined as follows. We have $2^N$ states; the 
{\it free-energy} of state $i$ is given by 
\begin{equation}
F_i = E_i - T S_i 
\label{F_i}
\end{equation}
where $E_i$ and $S_i$ are random independent variables, 
taken from a Gaussian
distribution of mean $0$ and standard deviation $\sigma_E=N^{0.5}/{\sqrt{2}}$
for the energy 
and $\sigma_S=N^{\alpha}/{\sqrt{2}}$ for the entropy, $\alpha$ being
an exponent that will play an essential role later. 
More explicitly, these probability distributions are
\begin{equation}
\rho_E(E_i)=\frac
{\exp \left({-\frac{E_i^2}{N}} \right) }
{\sqrt{N{\pi}}}
~~~
\rho_S(S_i)=\frac
{\exp \left({-\frac{S_i^2}{N^{2\alpha}}}\right)} 
{\sqrt{N^{2\alpha}{\pi}}}
\label{Probs}
\end{equation}
Note that for the $E_i$, the probability distribution is the same as in 
the REM. A simple way to visualize this model is to think to $2^N$ lines
in a temperature-free-energy
plot as in figure~\ref{fig_levelcross} where the intercept on the y axis 
is random as well as the slope of each line.

Being the sum of independent Gaussian random variables, the 
free-energy has a Gaussian distribution. Thus $F$ has zero 
mean and variance 
${{\sigma}_F}^2={{\sigma}_E}^2 + {{\sigma}_S}^2$. Explicitly, the variance 
of the free-energies satisfies
\begin{equation}
2{{\sigma}_F}^2 = N + T^2N^{2\alpha}
\label{Var_F}
\end{equation}

	If $\alpha=1/2$, the two terms contribute the same magnitude
at large $N$, 
while if $\alpha<1/2$, the variance is dominated by $\sigma_E$. 
The natural scaling then seems to be 
$\alpha=1/2$ in which case the ground state
energies and entropies are both extensive. In a T-F plot, we may expect 
these states to have crossings, 
like in the right part of figure~\ref{fig_levelcross}.

\subsection{The equivalent REM}

	Because of the Gaussian distribution of free-energies
at each temperature $T$, 
the model is equivalent to a REM whose energy variance is 
given by equation~\ref{Var_F}. This allows one to solve the 
thermodynamics of the model. The density of levels with free 
energy $F$ is 
\begin{equation}
{\rho}_F(F)=2^N \frac{e^{-\frac{F^2}{{2\sigma_F}^2} } }{\sigma_F\sqrt{2{\pi}}}
\label{NF}
\end{equation}
As expected, there is a critical dependence on $F$ for large $N$. 
If $|F| < F_0= {\sigma_F}{\sqrt{2N\ln{2}}}$, there is an exponentially large 
density of states and an extensive {\it configurational} entropy 
\begin{equation}
S_C(F)= N \ln{2} - \frac{F^2}{{2\sigma_F}^2}
\label{S_E}
\end{equation}
whereas if $|F|>F_0$, there are no levels at all in the thermodynamic limit 
and thus $S_C(F)=0$. Using the relation between configurational entropy
and temperature 
$T^{-1}={{\partial}S_C}/{{\partial}F}$, the critical 
temperature is  
\begin{equation}
T_c=\frac{\sigma_F}{\sqrt{2N\ln{2}}} 
\label{Tc_1}
\end{equation}
and thus $T_c$ satisfies the self-consistent equation
\begin{equation}
T_c=\frac{\sqrt{N + {T_c}^2N^{2\alpha}}}{2\sqrt{N\ln{2}}} 
=\sqrt{\frac{1 + {T_c}^2N^{2\alpha-1}}{4\ln{2}}}
\label{Tc_2}
\end{equation}

This lead us to distinguish three cases. First, if $\alpha<0.5$, then 
$T_c=1/ \(( 2\sqrt{\ln{2}} \))$, as in the original REM. There is a low 
temperature phase in which the partition function is dominated by a finite
number of states (those with the lowest free-energies); the 
entropy density in this phase is zero. There is also a high 
temperature phase for which
the number of states that contribute to the partition function is 
exponentially large in N. Second, for $\alpha=0.5$, one finds that 
$T_c=\frac{1}{\sqrt{4\ln{2}-1}}\approx 0.75$. The physics is the same 
as before, but the critical temperature is 
shifted. In both cases, we have a glassy model with a 
one-step replica symmetry 
breaking transition~\cite{Derrida81,GrossMezard84}. 
Third, if $\alpha>0.5$ one finds $T_c=+\infty$; there is no phase 
transition, and the systems remains in the ``low temperature phase'' 
at all temperatures. This can be considered to be unphysical because 
a microscopic Hamiltonian such as Eq~\ref{Hamil} will
always have a high temperature disordered phase. If we go back to
the Bray and Moore argument, in which 
entropy fluctuations grows as $l^{d_s/2}$,
we see that we should identify the Re-Rem parameter $\alpha$
with ${d_s}/(2d)$.
Happily, if one 
follows the prescription $\alpha={d_s}/(2d)$, 
there is no unphysical behavoir because $d_s\leq d$, so $\alpha \leq 0.5$.
	
Note that when $\alpha \leq 0.5$, 
the lowest free-energies scale linearly with $N$ and
the gap between the lowest free-energy
state and the first excitation is $O(1)$, as in the REM. Furthermore
if $\alpha<0.5$, the lowest free-energy density
does not change with $T$, whereas if $\alpha=0.5$, it grows as
$\sqrt{1+T^2}$. Finally in the case $\alpha>0.5$, 
the free-energy grows as $N^{\alpha + 0.5}$ and thus is
not extensive.

\subsection{Energy and entropy}

The lowest free-energy in the Re-Rem at each temperature is
easily found, and from it we can determine
the typical energy and 
entropy at each temperature. The lowest free-energy at each temperature 
is given by~\cite{Derrida81}
\begin{equation}
F_0 \approx - \sqrt{2N \log{2}} \sigma_f = - N \sqrt{\log{2} (1 + T^2 N^{2\alpha-1}) }
\label{F_min}
\end{equation}
The entropy in the low $T$ phase is
obtained from the 
derivative of this free-energy with respect to temperature:
\begin{equation}
S_0 \approx \frac {\sqrt{\log{2}}~T N^{2\alpha}}{\sqrt{1+T^2N^{2\alpha-1}}}
\label{S_min}
\end{equation}
which confirms that the entropy is extensive 
only for $\alpha=0.5$. 
To determine
the energy we use,
\begin{equation}
E_0 = F_0 + T S_0 
\approx -\frac {N\sqrt{\log{2}}}{\sqrt{1+T^2 N^{2\alpha-1}}}
\label{E_min}
\end{equation}
The energy is thus extensive if $\alpha\leq0.5$.

\section{Level crossings in the Re-Rem}
\label{sect_cross}

	Given our model with many states and a spin glass phase, we 
focus on the problem of level crossings. We have $2^N$ levels and 
randomness in entropy and 
energy. We want to 
follow the lowest free-energy state and see if this 
state changes with temperature, and how often it does if so. 
	First, we will provide some analytical constraints, deriving a 
scaling law governing the number of crossings in a $\delta T$ interval; 
then we will determine the corresponding 
scaling function by a numerical computation.
	
\subsection{Level crossings via thermodynamics}

	At $T=0$, the lowest level is the one with the minimum energy, entropy 
plays no role. If one follows the line starting from this minimum 
energy in temperature, we can expect some other levels, 
with larger energies but also more negative entropies, to cross this line. 
This is what happened in the Bray and Moore analysis, showing that
zero temperature chaos is inevitable. But what happens at $T>0$? Does the lowest free-energy state change at each temperature? What is the density of crossings? The problem of knowing which level is lowest at each temperature and 
of computing crossing statistics seems rather non-trivial; 
however thanks to the mapping to the REM, 
it is possible to compute the temperature dependence of the best 
level, and to deduce from that many properties of the crossings.
	
	Let us begin with some remarks. First, the role of 
the exponent $\alpha$ is just to rescale the temperature definition by a factor
$N^{\alpha}$ as is evident from equation~3. Thus if one
solves the problem for $\alpha=0$, changing to an $\alpha \neq 0$ 
just divides each crossing temperature 
by $N^{\alpha}$. As a consequence, 
the total number of crossings does
not depend on $\alpha$. Second, this total number of crossings 
(the number of times the lowest free-energy level changes its slope) is at 
most of $O(N)$. 
This can be shown using the following argument. Take 
all $2^N$ states and consider the lowest one $\mathcal L_1$ at $T=0$. 
The entropy is random and uncorrelated with the energy, so if you
take the next state $\mathcal L_2$ in energy, you will have a good 
chance ($50{\%}$) that
it will have a more negative entropy and thus will cross $\mathcal L_1$.
Suppose this is the case. 
A further state $\mathcal L_i$ with a still greater 
energy will cross $\mathcal L_2$ only if its slope is more 
negative than that of $\mathcal L_2$. The entropy 
of a successful candidate $\mathcal L_i$ must be the lowest 
of $i$ entropies taken at random. Since by hypothesis
$\mathcal L_2$ has the lowest entropy of $2$ entropies
taken at random, we have to let $i$ go up to $4$ to have
a $50{\%}$ chance of finding a state that will cross 
$\mathcal L_2$. Extending this argument, to get the $k^{th}$ crossing  
one will need to examine $O(2^{k})$ successive levels.
Furthermore, it is not difficult to show that
when going from $\mathcal L_{2^k}$ to $\mathcal L_{2^{k+1}}$
the energy increases by $O(1)$.

\begin{figure}
\centerline{\hbox{\epsfig{figure=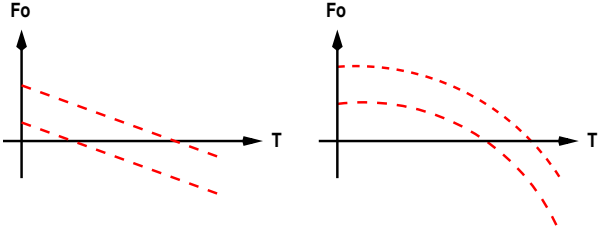,width=8cm}}}
\caption{The lowest states in the Re-Rem remain inside a ``tube''.
Left: $\alpha>0.5$, the dependence is linear in $T$;
right: $\alpha\leq0.5$ the tube is curved.}
\label{fig_tube}
\end{figure}

Of course, these $O(N)$ crossings are 
not the end of the story because some of them 
may be irrelevant. Take for example the right part of 
figure~\ref{fig_levelcross} 
where the third energy state crosses the second one, 
but where the fourth one crosses the second one {\it before} the third does.
From this simple example we see that not all crossings count;
we can only say that the number of levels 
that become at some point the state of lowest free-energy is at 
{\it most} $O(N)$, and that the energy at each level crossing increases
by at {\it least} $O(1)$.

In the large $N$ limit, a scaling behavior sets in. But before 
being quatitative, let us see what qualitative prperties imply
level crossings. The starting point is the fact that, at fixed $T$, 
the Re-Rem reduces to a REM. The lowest free-energy at each temperature 
scales as in equation~\ref{F_min} and its sample to sample fluctuations 
are of $O(1)$. Thus the lowest state resides in a ``tube''
(see figure~2) in the (T,F) diagram.
Clearly, if the tube is not linear in temperature,
there must be some level crossings. 
We will now see how this imposes strong constraints on the crossing statistics.

	First, consider the case where {\bf $\alpha>0.5$}. At large $N$, we get
\begin{equation}
F_0 \propto - N^{\alpha+0.5} T 
\label{F_min_bigalpha}
\end{equation}
The linear temperature dependence suggests that 
the tube has one, or just of few states passing through
it for all $T$. 
Moreover the slope of this tube is of $O(N^{\alpha+0.5})$ 
which is the scale for the lowest entropy value obtained by 
taking $2^N$ values in a Gaussian of variance $N^{2\alpha}$. 
So if at $T=0$, the best state is the one with lowest $E$, at any 
strictly positive 
temperature the best state is the one with the best slope. 
This means that all level crossings take place at zero temperature;
there is zero temperature chaos but no chaos
at $T>0$!

When {\bf $\alpha \leq 0.5$}, we have 
\begin{equation}
F_0 \approx - N \sqrt{\ln{2}} - \frac{N^{2\alpha} \sqrt{\ln{2}} T^2}{2}
\label{F_min_moderatealpha}
\end{equation}
	Now, the dependence in $T$ is non-linear; there must
be some crossings at $T>0$ because the lowest energy state changes its 
slope in temperature and in $N$. This suggest that the number of crossings
in a $\delta T$ interval diverges with $N$ for every temperature.

	Finally, if {\bf $\alpha=0$}, we obtain 
\begin{equation}
F_0 \approx - N \sqrt{\ln{2}} - \sqrt{\ln{2}} T^2
\label{F_min_alpha0}
\end{equation}
so the dependence on $T$ is $N$-independent.
Since the energy increase of successive crossings is at least of $O(1)$, 
it is not possible to pass an infinite number of levels in the tube. 
Thus there should only be a finite number of crossings 
in each $\delta T$ interval. 

Now we also know that changing $\alpha$ is the same as re-scaling the 
temperature; this gives a strong constraint on the number of crossings in 
a given interval. Let's assume that the 
scaling of the mean number of crossings in 
a temperature interval $\delta T$ is of the form
\begin{equation}
{\cal{N}}_N(T,T+\delta T)= N^{\gamma(\alpha)} \delta T 
g \((
\frac {T}{N^{\kappa(\alpha)}} \))
\label{scaling_form_1}
\end{equation}
Then the invariance under the transformation
$\alpha_1 \rightarrow \alpha_2$ and $T \rightarrow  T N^{\alpha_1-\alpha_2}$ gives
\begin{eqnarray}
\begin{array}{c} 
\gamma(\alpha_1)-\gamma(\alpha_2)=\alpha_1-\alpha_2\\
\kappa(\alpha_1)-\kappa(\alpha_2)=\alpha_2-\alpha_1\\
\end{array}
\label{Demo_scal1}
\end{eqnarray}
It follows that
\begin{eqnarray}
\begin{array}{c} 
\gamma(\alpha)=\alpha+\beta_1\\
\kappa(\alpha)=-\alpha+\beta_2\\
\end{array}
\label{Demo_scal2}
\end{eqnarray}
Furthermore, we must respect the constraints derived previously.
First, assuming $g(x)$ decays quickly (not as a power law) at 
large $x$, the disappearance of $T>0$ crossings when $\alpha>0.5$
gives $\beta_2=0.5$. 
Second, since $\alpha=0$ gives crossing density at large $N$,  
we must have $\beta_1=0$.
The
scaling form for the mean number of crossings in a temperature 
interval $\delta T$ is then
\begin{equation}
{\cal{N}}_N(T,T+\delta T)= N^{\alpha} \delta T 
g \(( \frac {T}{N^{0.5-\alpha}} \))
\label{scaling_form}
\end{equation}
From this formula, we deduce
that the total number of crossings is in fact $O(\sqrt{N})$,
much less than the bound $O(N)$. Also if $g(0)$ is finite,
the case $\alpha=0$ leads to a mean temperature of
the $k^{th}$ crossing that is proportional to $k$.

\subsection{Numerical investigation }

It is of some use to illustrate and check these results by a numerical 
procedure but we also want to compute the function $g(x)$. A simple way 
to do this is to generate $2^N$ random energies, $2^N$ random entropies,
and to check for crossings, but such a procedure require a computation time
that is exponential in $N$.
Instead we used a method which allows us to have an 
algorithm linear in $N$, and thus to simulate up to $2^{800}$ levels. 
With no loss of generality we can focus on the case $\alpha=0$;
we have simulated $10000$ instances for $N=100,200,300,400,600$ and $800$.

\begin{figure}
\centerline{\hbox{\epsfig{figure=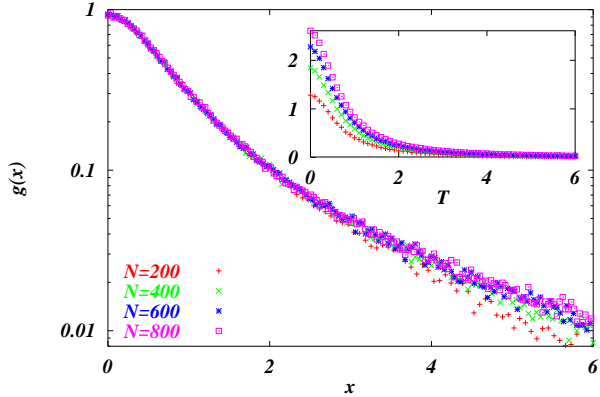,width=8cm}}}
\caption{Scaling function $g(x)$ in the Re-Rem. 
The inset shows the raw data for the mean number of crossings
versus temperature for $\alpha=0.5$.}
\label{fig_reremcross1}
\end{figure}

We explain our algorithm in the appendix for readers interested
in algorithmic issues. Basically, our method constructs 
in a time proportional to $N$ the series of 
crossing points for a random realization of the Re-Rem.
Then we measure the statistics of these crossings by averaging over
many instances, repeating this for different values of $N$.
The scaling function $g(x)$ is then obtained by 
using equation~\ref{scaling_form}; the results
are shown in figure~\ref{fig_reremcross1}. 
The data collapse is excellent, showing
that the $T/ \sqrt{N}$ scaling works very well. We
do see some deviations at small $N$ in the tail; these
correspond to corrections to scaling, and are small (invisible)
beyond $N=500$.

From this numerical computation of $g(x)$, we can extract the number
of crossings in any temperature interval for any
$\alpha$ and $N$.
Consider first the case $\alpha=0.5$ which is the appropriate
value in the mean field picture. One has
\begin{equation}
{\cal{N}}^{\alpha = 0.5}_N (T,T+\delta T) = \sqrt{N}\delta T g(T)
\label{scaling_form_test}
\end{equation}
Focusing on the interval
${T_c}/{3}$ to ${2T_c}/{3}$, we find
about $6\sqrt{N / 1000}$ crossings, that
is $6$ crossings for $N=1000$.

One can also consider other values for 
the exponent $\alpha$. In the framework of 
the EA model in $d=3$, the fractal dimension of
the surface of system-size domains has been 
measured~\cite{KrzakalaMartin00,PalassiniYoung00a},
leading to $\alpha=d_s/2d \approx 0.46$.
For this value of the exponent the mean number of crossings
in the ${T_c}/{3}$ to ${2T_c}/{3}$
interval is now reduced to about $3$ when $N=1000$.

The main conclusion
of this section is that the number of crossings in a temperature 
interval grows as
a power of $N$ if $0 < \alpha \le 0.5$ but with a small {\it prefactor}. 
Thus, if we simply map our Re-Rem
results to estimate the strength of chaos that should arise in
Monte Carlo simulations, we find that there will be very
little sign of chaos
until $N$ is at least a few hundred.
However, the mapping from the Re-Rem
most certainly over-estimates significantly the number
of crossings in microscopic models (cf. the next
section and the final discussion).

\section{Droplet extension of the Re-Rem}
\label{sect_droplet}

As given so far, our Re-Rem has excitation energies of $O(1)$.
However, it is possible to modify the model in order to mimic the 
{\it droplet/scaling}~\cite{BrayMoore86,FisherHuse88} picture of spin 
glasses. There 
low-lying excitation energies are $O(N^{{\theta}/d})$,
that is $O(L^{\theta})$ where $\theta$ is the stiffness exponent
which is at the
heart of the droplet and scaling pictures of
spin glasses.

\subsection{REM revisited: The Scaling-REM}

To construct the Scaling-REM (S-REM), we note that
it is not appropriate to identify 
single spin configurations with states, a valley containing
a large number of similar configurations. Thus
we take a number of states that is much smaller that $2^N$; as a 
consequence, the variance of their energy will no longer be $O(N)$.
Our method is thus to rescale the energies 
by a factor $N^{{\theta}/{(2d)}}$ while the number of states 
is reduced
from $2^N$ to ${2^N}^{\beta}$ (with $\beta<1$). This gives
\begin{equation}
2{{\sigma}_E}^2 = N^{1+\frac{\theta}{d}}
\end{equation}
\begin{equation}
S_C(E)= N^{\beta} \ln{2} - \frac{E^2}{{2\sigma_E}^2}
\end{equation}
Then $T_c=\sigma_E / \sqrt{2N^{\beta}\ln{2}}$ and 
$E_0= -{\sigma_E}{\sqrt{2N^{\beta}\ln{2}}}$. 
We want an extensive ground state energy at $T=0$; thus
\begin{equation}
E_0 = - {\sqrt{N^{\beta}\ln{2}}} \sqrt{N^{1+{\theta}/{d}}}
\label{S_E0_drop}
\end{equation}
imposes $\beta=1-\frac{\theta}{d}$. 
It follows that the critical temperature is 
now given by $T_c=N^{\theta/d} / 2\sqrt{\ln{2}}$, 
and so $T_c > 0$ only
if $\theta \geq 0$. Note that for $\theta > 0$, $T_c \to \infty$
as in the Directed Polymer in a Random Medium (DPRM) in $1+1$ dimension, 
in contrast 
to what happens in realistic spin glasses. Actually, this S-REM
is very close to the DPRM. Consider for instance the 
exponents giving (i) the finite size corrections to
the ground state energy, (ii) the 
sample to sample fluctuations and (iii) the energy of the 
first excited state; they are the same 
(i.e., $\theta / d$) in the two models, whereas
two distinct exponents arise in the droplet 
model~\cite{BouchaudKrzakala02}. Nevertheless,
this S-REM provides an interesting picture of the 
low temperature phase that is quite
compatible with the scaling/droplet picture there,
even if it misses $T_c < \infty$.

\subsection{Re-Rem revisited}

We can perform the analogous scaling extension
for the Re-Rem and obtain a Scaling-Re-Rem where
low-lying excitation energies scale as $N^{\theta/d}$. All 
the computations are straight-forward.
One finds that the critical temperature of the S-Re-Rem is given by 
\begin{equation}
T_c=\frac{\sqrt{N^{1+{\theta}/{d}} + T^2N^{2\alpha}}}{2\sqrt{N^{1-{\theta}/{d}}\ln{2}}}=\frac{N^{{\theta}/{d}}\sqrt{1+T^2 N ^{2\alpha-{\theta}/{d}-1}}}{2\sqrt{\ln{2}}}
\end {equation}
and that 
$F_0 \propto - N \sqrt{1+T^2N^{2\alpha-1-{\theta}/{d}}}$. Now 
letting ${\alpha}'=\alpha-\theta/(2d)$, we have
\begin{equation}
T_c= \frac{N^{{\theta}/{d}}\sqrt{1+T^2 N ^{2{\alpha}'-1}}}{2\sqrt{\ln{2}}}
\label{Tc_drop_fin}
\end{equation}
\begin{equation}
F_0 \propto - N \sqrt{1+T^2N^{2{\alpha}'-1}}
\label{S_F0_drop_fin}
\end{equation}
From this, we see that the thermodynamics is the same as that of 
the Re-Rem but where $\alpha$ has been replaced by 
$\alpha-{\theta}/{(2d)}$. We can also apply this correspondence
to the formulae for the
level crossing properties~\footnote{Note that 
the width of the tube now grows as $N^{\theta/d}$.},
again again that there is temperature chaos. More precisely, 
if we start from the scaling form~\ref{scaling_form_1} then using 
${\alpha}'$ instead of $\alpha$, 
equation~\ref{S_F0_drop_fin} implies as before that $\beta_2=0.5$, but
$\beta_1 \neq 0$ since
the width of the tube grows as $N^{\theta/d}$. However, the number
of levels being $2^{N^{\beta}}$, the total number of crossing goes
as $\sqrt{N^{\beta}}$ and the general scaling for $\theta \neq 0$ is
\begin{equation}
{\cal{N}}_N(T,T+\delta T)= N^{\alpha-{\theta}/{d}} \delta T 
g \((
\frac {T}{N^{0.5-\alpha+{\theta}/{2d}}}
\))
\label{scaling_form_theta}
\end{equation}

\subsection{Criterion for temperature chaos}

Even if the S-Re-Rem model is a bit ad-hoc,
it may correctly describe some glassy systems at low temperature. 
The general criterion for temperature chaos 
depends only on the value of parameters $\alpha$ and $\theta/d$. 
If $\alpha-\theta/d<0$, there is no chaos at all, all crossings being 
pushed to $T=\infty$. If $\alpha-\theta/d=0$, the density of crossings is 
finite; this is what we call {\it weak temperature chaos}. 
If $0 < \alpha-\theta/d$ while $ \alpha-\theta/(2d) \leq 0.5$, 
there are an infinite number of crossings in each temperature
interval, that is 
{\it strong temperature chaos}. Finally, 
if $\alpha-\theta/(2d) >0.5$, there is {\it no finite temperature chaos} 
because all crossings take place at $T=0$. 
Note that if one accepts $\alpha=d_s/2d$, we recover the original
Bray and Moore argument.
However, within our model we can estimate the number
of crossings, and comparing to the discussion of section~\ref{sect_cross},
the case $\theta>0$ leads to even fewer crossings than when
$\theta=0$.

\section{Temperature chaos in the na\"{\i}ve mean field framework}
\label{sect_chaos}

Ideally, one would like to confront the Re-Rem predictions with
the chaos properties arising in a realistic model. This requires
defining finite volume states or valleys
and following them as a function of temperature.
A natural approach is to use the TAP~\cite{ThoulessAnderson77} 
equations where each state is parametrized by
its magnetizations $m_i = \langle S_i \rangle$. 
However, these equations have 2 serious drawbacks:
(i) on the $3$-dimensional lattice, the TAP equations fail to
give non-paramagnetic solutions~\cite{LingBowman83}; (ii) 
even for the SK model~\cite{NemotoTakayama85},
having $N< \infty$ (which is unavoidable numerically)
leads to unsurmountable convergence problems~\cite{Nemoto87}.
The source of these difficulties
seems to be the strength of the retro-action term. An ad-hoc
solution consists in either
reducing or neglecting this term. Following previous 
authors~\cite{LingBowman83,MuletPagnani01}, we
choose the second option which corresponds
to the na\"\i ve mean field equations (NMFE)~\cite{BraySompolinsky86}.

\subsection{Na\"{\i}ve Mean Field Equations }

The NMFE are
\begin{equation}
m_i= \tanh{({\beta}\sum_{j} J_{ij} m_j)}
\end{equation}
The solutions to these equations define the na\"{\i}ve mean field (NFM) 
states; they 
correspond to the stationary points of the NMF free-energy
functional
\begin{equation}
F(\{ m_i \}) = E(\{ m_i \}) - T S(\{ m_i \})
\end{equation}
where the energy and the entropy functionals are given by
\begin{equation}
E= - \sum_{<ij>} J_{ij} m_i m_j
\end{equation}
\begin{equation}
S= -\sum_i \frac{(1+m_i)\ln(1+m_i)+(1-m_i)\ln(1-m_i)}{2}
\end{equation}
$F$ has both local minima and saddles; we follow
the TAP approach and construct in our numerics only local minima.

Our procedure is as follows. For each disorder instance,
we will produce two NMF states
$\mathcal L_1$ and $\mathcal L_2$ and will work out
their dependence on temperature. We start by building their
$T=0$ intercept: $\mathcal L_1$ is in fact the ground state,
while $\mathcal L_2$ is a large-scale low-energy excitation.
(To obtain the ground state, we use
a state-of-art algorithm~\cite{HoudayerMartin01}. In virtue of the method 
used to generate excitations~\cite{KrzakalaMartin00}, the 
excitation energies are 
$O(1)$ and do {\it not} grow when increasing $N$.)
We then numerically determine how the $\mathcal L_i$ evolve as
we increase the temperature from $0$ to $T_c$. 
In practice, we increase the temperature $T$ in sufficiently small 
steps ($\Delta T = 0.05$) and perform
steepest descent on the free-energy functional, using
as starting point $\mathcal L_i$ at the previous temperature.
With this procedure, referred to as {\it heating} of the ground state 
in~\cite{MuletPagnani01}, we consider two NMF states
that evolve with increasing temperature.

\subsection{Numerical results for the $3-d$ EA model}

	\begin{figure}
 \resizebox{0.45\textwidth}{!}{
 \includegraphics{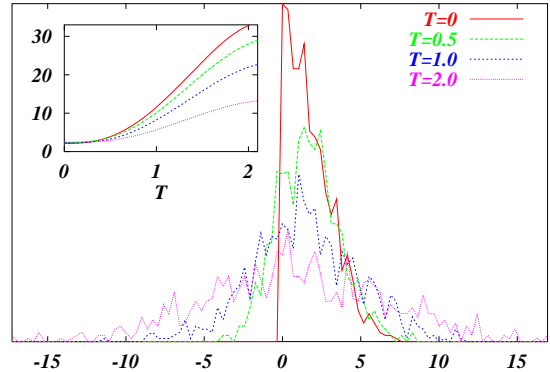}}
 \caption{$P(\delta F)$ at different temperatures 
for $0 \leq |q| \leq 0.05$ excitations in L=12 EA model. Inset: the variance
of $\delta F$ for $L=8,10,11$ and $12$ (from bottom to top).}
\label{EA_CW}
\end{figure}

First, let us give some qualitative properties of this NMF system.
We find a non-zero Edwards-Anderson order
parameter $q_{EA}={\sum_i {m_i}^2}/{N}$ for any instance and 
any state when $T<T_c$ ($T_c{\approx}5$
in the $3-d$ NMF EA model, and $T_c{\approx}2$ in the NMF SK model), and 
$q_{EA}$ is self-averaging. If 
we follow a given state, 
the two temperature spin overlap 
$q_{{0}{T}}={\sum_i {m_i}^{(T=0)}{m_i}^{(T)}}/{N}$ 
decreases and reaches zero when $T=T_c$. Furthermore, 
the equal temperature mutual overlap between $\mathcal L_1$
and $\mathcal L_2$, 
$q^{{\alpha}{\beta}}_{TT}={\sum_i {m_i}^{\alpha}(T){m_i}^{\beta}(T)}/{N}$,
also decreases smoothly
when going from $T=0$ to $T_c$, suggesting that evolving those 
states according to the NMF equations 
mainly just reduces the size of the $m_i=<S_i>$.
Do we get level crossings as would be expected
if there was temperature chaos? We are particularly interested in the 
behavior of state to state free-energy fluctuations at very low temperatures
as this quantity is at the heart of our Re-Rem picture.

For this model, we have used on average $1000$ ground states
and excitations for sizes $L=8,10,11$ and $12$. We focus in
all that follows on the instances where our excitations
have an overlap with the ground state satisfying
$0 \le |q| \le 0.05$. However
the data for other
overlap intervals
have a qualitatively similar behavior.

According to zero-temperature chaos, when going from $T=0$ to $T>0$, 
the valley $\mathcal L_1$ 
should be above the valley $\mathcal L_2$ with a strictly postive
probability when $N$ grows. We find this to be the case, and
have found that the probability
to find such an inversion 
increases with temperature. 
To understand this more quantitatively,
define $\delta F$ to be the free-energy difference
$F(\mathcal {L}_2) -  F(\mathcal {L}_1)$.
Figure~\ref{EA_CW} shows the distribution $P(\delta F)$ 
at different temperatures for $L=12$.
The variance of this distribution increases with temperature.
The curves for different $L$ and different window 
overlaps look similar, so  
once $T>0$, there is a positive probability
that $\delta F < 0$, corresponding to a level crossing.
Since there are an infinite 
number of such excitations in the 
thermodynamics limit, and since they plausibly all behave the same way,
these crossings suggest that there is indeed temperature chaos.

To make contact with our Re-Rem picture, one should extract $\alpha$ from
the $L$ dependence of the variance of $\delta F$. If $\alpha >0$, the 
variance of
$\delta F$ will diverge at 
large $L$ as $L^{2\alpha d}$. 
In the inset of figure~\ref{EA_CW} we plot the variance of $\delta F$ at 
different temperatures. At any given temperature, we can fit
its L dependence; not surprizingly, just as was previously found
at $T=0$~\cite{KrzakalaMartin00,PalassiniYoung00a},
the data can be fit with $\alpha=0.5$ (the mean field 
case $d=d_s$) while the best fit gives $\alpha \approx 0.46$ 
($d_s \approx 2.76$). What is important is that $\alpha$ is certainly 
not 0. This suggests that this system have strong chaos.

\subsection{Numerical results for the SK model}
	
Recently, the na\"{\i}ve mean field
framework was used by Mulet {\it et al.}~\cite{MuletPagnani01} 
to study temperature chaos in the SK model. Although they
concluded that there is no chaos, we feel it is necessary
to reconsider that conclusion in view of the Re-Rem analysis.

Compared to the EA model, the SK is numerically difficult because
of its long range interactions. More numerical effort is needed
to find ground states or excitations and also for doing 
the steepest descent on the NMF free-energy functional.
We were thus limited to ``small'' values of $N$:
we used $N=64, 128$ and $192$, performing statistics
using $50$ ground states
and excitations for each value of $N$.

\begin{figure}
 \resizebox{0.45\textwidth}{!}{
 \includegraphics{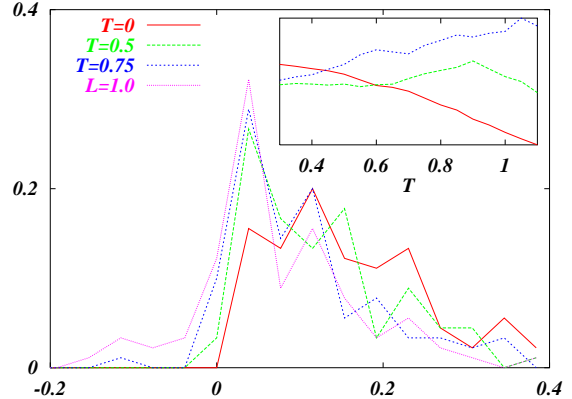}}
 \caption{$P(\delta F)$ at different temperatures 
for $0\le |q| \le 0.05$ excitations in the SK model ($N=192$). Inset:
the variance of $\delta F$ for $L=64,128$ and $192$ as a function 
of $T$ (bottom to top).}
\label{SK_CW}
\end{figure}

Our conclusions are qualitatively similar
to those we obtained for the $3-d$ EA model (see figure~\ref{SK_CW}).
We find that there is a strictly positive probability that $\mathcal {L}_1$
and $\mathcal {L}_2$ cross; fits of the free-energy
fluctuations give a value for $\alpha$ fully compatible with $0.5$.
These results are those that are expected in the Re-Rem picture;
it is thus appropriate to conclude that there {\it should be}
temperature chaos. However, note that the variance
of $\delta F$ grows only slowly with $N$ so very few crossings are expected
until much larger values of $N$ are reached.

\section{Discussion}
\label{sect_discussion}

This work has two main points.
First, we have developped a solvable model of spin glasses inspired from
the REM that allows a precise study of temperature chaos. This 
model has strong temperature chaos unless its exponent
$\alpha$ is artificially set to zero. However; this chaos is absent until
the system size $N$ reaches about several hundred.
Second, we have argued that this Re-Rem model should
be relevant for real spin
glasses by considering the particular case of na\"{\i}ve mean field 
equations.
The numerical study of that system turned out to be in complete
agreement with the expectations of the Re-Rem picture,  
giving strong evidence that temperature chaos does occur
in spin glass models such as the $3-d$ Edwards-Anderson and the SK models.

The question of temperature chaos has lingered on for many
years because of lack of numerical evidence.
One may considering the absence of chaos signals in past and present MC
simulations by the fact that in the Re-Rem, $N$ must be quite large
for chaos to appear. 
Indeed, we saw that in the Re-Rem chaos becomes clear for $N \simeq 1000$ 
but it is clearly inappropriate to consider that one has
$2^N$ states in a physical system of $N$ spins. Having far
fewer states will lead to fewer crossings, and so it may be 
that no chaos will be seen from MC simulation in the near future.
However, we believe that our na\"{\i}ve mean field study suggests on
the contrary that chaos in realistic systems
is within grasp of current Monte Carlo techniques.
One way to do this is to consider for instance the
$3-d$ EA model and to extract at each temperature the
states or ``valleys'' by clustering the
equilibrium configurations. Then following these
states in temperature will allow one to compare directly
with the Re-Rem picture.

In the future, it would be interesting to extend the Re-Rem
to more complex systems such as the Generalized Random-Energy-Model. 
In such a G-Re-Rem model, one would be able to understand the effects of 
2-step, 3-step and continuous RSB. Another interesting possibility 
is to study the Re-Rem with more realistic behavior than a pure linear 
dependence in $T$, the free-energies of levels being a more complicated 
function of $T$.

\vspace{1cm}

\centerline{{\bf Acknowledgments}}
We wish to thank M.~M\'ezard, G.~Parisi and M. Sasaki for 
stimulating discussions. 
F.K. acknowledges a fellowship from the
MENRT. The LPTMS laboratory is an Unit\'e de 
Recherche de l'Universit\'e Paris~XI 
associ\'ee au CNRS.

\vspace{1cm}
\centerline{{\bf Appendix}}

\bigskip

Here we explain in detail our algorithm. We want to construct efficiently 
the series of crossing points
for a realization of the Re-Rem
having $2^N$ states or levels, level $i$ having
energy $E_i$ and entropy $S_i$. It is convenient to assume
that these states have been ordered in increasing value of their
energies. We will start with the ground state: this level $l(0)=0$ has
energy-entropy $(E_{l(0)},S_{l(0)})$. Then we 
will continue by finding the next level
$l(1)$ that has a lower entropy than that of $l(0)$, 
i.e., $S_{l(1)} < S_{l(0)}$. This construction is extended
by recurrence: given the level number $l(n)$ and values
($E_{l(n)},S_{l(n)}$), we will generate with the correct
probability distribution of the next level
$l(n+1)$ that has $S_{l(n+1)} < S_{l(n)}$ as well as
the quantities $(E_{l(n+1)},S_{l(n+1)})$.
The key point is that all these quantities
can be generated recursively without
``looking'' at all $2^N$ states. Once this list of
($l(n),E_{l(n)},S_{l(n)}$) values is constructed
(this takes $O(N)$ operations), we can
then reconstruct the lowest state at any temperature,
and thus the statistics of the crossings.

To begin our algorithm, we set $n=0$, $l(0)=0$ and choose
$S_{l(0)}$ according to its Gaussian distribution. Then 
we must generate $E_{l(0)}$ (note that $E_{l(0)}$ and
$S_{l(0)}$ are uncorrelated random variables). The probability density 
of $E_{l(0)}$ is
\begin{equation}
P(E_{l(0)}) = 2^N {\rho}_N(E_{l(0)}) {[1-Q(E_{l(0)}]}^{(2^N-1)}
\label{algo_E0}
\end{equation}
where ${\rho}_N(E)$ is the probability density 
of energies in the Re-Rem model (a Gaussian) and $Q$ is defined by 
\begin{equation}
Q(E) = \int_{-\infty}^{E}  {\rho}_N(x) dx
\label{algo_Q}
\end{equation}
To generate $E_{l(0)}$ with its distribution, we use the accept/reject method.
Finally we set $E_{old}=E_{l(0)}$ and $S_{old}=S_{l(0)}$.

Now we enter the recursion. Given the values
$l(n)$, $E_{old}$ and $S_{old}$, we first determine
the next level $l(n+1)$ that has an entropy 
below $S_{old}$. Let $\Delta l$ be defined by
$l(n+1) = l(n) + \Delta l$;
$\Delta l$ is exactly the
number of times one has to generate 
a random entropy $S_{new}$ according to the Re-Rem entropy 
distribution until we find a value satisfying 
$S_{new} < S_{old}$. Define $p$ to be the 
probability to get a slope smaller than $S_{old}$ with {\it one} trial; then
\begin{equation}
p=\int_{-\infty}^{S_{old}}  {\rho}_N(S) dS
\label{algo_Sprime}
\end{equation}
Clearly, $\Delta l$ (the number of attempts made before 
reaching an entropy value lower than $S_{old}$)
is a random variable; its probability distribution is
\begin{equation}
P(\Delta l)=(1-p)^{\Delta l -1}p
\label{algo_S}
\end{equation}
We thus generate $\Delta l$ according to this distribution.
This gives us the new level to consider,
$l(n+1) = l(n) + \Delta l$. If $l(n+1) \geq 2^N$, we stop
as such a level does not exist.
Otherwise, $l(n+1)$ is a legitimate level, and we proceed
by determining ($E_{l(n+1)},S_{l(n+1)}$). 

The {\it entropy} of 
level $l(n+1)$ is easy to generate because its distribution
is that of any $S_i$ subject to the condition that
$S_{new}<S_{old}$. The only difficulty is for generating $E_{l(n+1)}$.
To do that, start with
$P(E_{old},E_{new})=P(E_{old})P(E_{new}|E_{old})$ where
the quantities $l(n)$ and $l(n+1)$ are considered as known. 
From this relation, we derive
\begin{eqnarray}
P(E_{new}|E_{old})=
\left(\begin{array}{c} 
l(n+1)-l(n)-1
\\ 
M-(l(n)+1)
\end{array}\right)
\nonumber \\
\times ~{Q(
{E_{old},E_{new}}
)}^{\Delta l -1}
(M-l(n+1))
{\rho}_N(E_{l(n+1)})
\nonumber \\
\times ~\frac
{(1-Q(E_{l(n+1)}))^{M-(l(n+1)+1)}}
{(1-Q(E_{l(n)}))^{M-(l(n)+1)}}
\label{algo_En}
\end{eqnarray}
when $M=2^N$ and $Q(a,b)=Q(b)-Q(a)$. We sample this
distribution by the accept-reject method again. 

Equation~35	has a simple interpretation; we have built a set of ``volumes'' in
the space of $E_i$s. The relative space we can explore now has weight 
${[1-Q(E_{l(n)})]}^{M-(l(n)+1)}$. In this space, we first generate
$\Delta l -1$ energies in the interval $[E_{l(n)},E_{l(n-1)}]$ and 
this gives the first part. Then we assign the level whose energy is
$E_{l(n+1)}$. 
There are $M-l(n+1)$ ways to choose the level
among the remaining ones. Finally,
the last $(M-(l(n+1)+1))$ energies have to be assigned and then we
must normalize the distribution. Given this $E_{new}$, we set
$E_{l(n+1)}$ to $E_{new}$ and we are finished. To continue the
recursion, we set $E_{old}$ to $E_{new}$, $S_{old}$ to $S_{new}$
and $n$ to $n+1$.

To summarize the whole procedure,
using these probability distributions, we can directly create an 
instance of a list of crossings 
by producing the $O(N)$ $E_{l(n)}$ and $S_{l(n)}$ that will cross and 
then study these crossings (this is a simple problem). 
From a numerical point of view however, these distributions 
are not so easy to compute, and one has to take care of
machine rounding errors; that source of error limited us to $N < 1000$.

\vspace{1cm}

											
%

\bibliographystyle{prsty}
\bibliography{../../Bib/references}

\end{document}